# From Symbol to Meaning: Ontological and Philosophical Reflections on Large Language Models in Information Systems Engineering

José Palazzo Moreira de Oliveira[1] [0000-0002-9166-8801]

[1] Federal University of RGS, Instituto de Informática, Porto Alegre, Brazil
`palazzo@inf.ufrgs.br`

**Abstract.** The advent of Large Language Models (LLMs) represents a turning point in the theoretical foundations of Information Systems Engineering. Beyond their technical significance, LLMs challenge the ontological, epistemological, and semiotic assumptions that have long structured our understanding of information, representation, and knowledge. This article proposes an integrative reflection on how LLMs reconfigure the relationships among language, meaning, and system design, suggesting that their emergence demands a re-examination of the conceptual foundations of contemporary information systems. Sketching on philosophical traditions from Peirce to Heidegger and Floridi, we investigate how the logic of generative models both extends and destabilises classical notions of ontology and signification. The discussion emphasises the necessity of grounding LLM-based systems in transparent, ethically coherent frameworks that respect the integrity of human-centred knowledge processes. Ultimately, the paper argues that LLMs should be understood not merely as tools for automation but as epistemic agents that reshape the philosophical and semiotic foundations of information systems engineering.

## 1 Introduction

The title From Symbol to Meaning: Ontological and Philosophical Reflections on Large Language Models in Information Systems Engineering emphasises the transition from symbol to meaning by drawing on the perspective of Semiotics, in which a symbol is understood as a sign that represents something within a system of interpretation rather than possessing inherent meaning. In this sense, large language models operate primarily on symbolic structures, tokens, words, and syntactic patterns, while the notion of meaning emerges only when these symbols are interpreted within a conceptual and ontological context. The title, therefore, signals a reflection on how computational systems manipulate symbolic representations and how these representations relate to semantic understanding and domain knowledge within the foundations of Information Systems Engineering.

The field of Information Systems Engineering (ISE) has undergone profound transformations since its inception, shaped by both academic inquiry and practical experience. Originating from the need to address complex organisational problems, ISE emerged as a response to the challenges of managing resources, processes, and communication within increasingly industrialised societies. Early urban centres already required coordinated systems for logistics, storage, and security, laying the conceptual foundations for modern information systems.

The ethnographic dimension of ISE, crucial for understanding user contexts, has its intellectual roots in the work of Bronisław Malinowski (Malinowski 1929), who introduced the concept of participant observation in the early twentieth century. His emphasis on immersive fieldwork established the methodological principle that a comprehensive understanding of a system requires deep engagement with the environment in which it operates. This insight remains central to ISE practice: effective system design depends on an interpretative understanding of human behaviour and organisational culture.

Within this broader perspective, the ethnographic dimension of Information Systems Engineering illustrates how the field can benefit from intellectual exchanges across multiple areas of knowledge, thereby preventing the formation of isolated disciplinary silos. The ethnographic approach, rooted in Bronisław Malinowski's concept of participant observation, demonstrates that understanding information systems requires not only technical or formal modelling perspectives but also interpretative insights into human practices and organisational contexts. By incorporating perspectives from Ethnography and related social sciences, researchers from different domains can contribute complementary viewpoints to the study and design of systems. Such interdisciplinary engagement fosters a more integrated understanding of technological artefacts, human behaviour, and institutional environments, encouraging collaboration rather than fragmentation within the broader research landscape.

In parallel, the systemic dimension of this interpretative approach is grounded in the work of Ludwig von Bertalanffy (Von Bertalanffy 1972), one of the principal architects of General Systems Theory. His vision extended beyond the boundaries of biology, proposing a holistic framework that could integrate knowledge across diverse fields, including cybernetics, psychology, sociology, and education. Within the context of Information Systems Engineering, this perspective reinforces the idea that systems cannot be understood in isolation but only through the dynamic interrelations among their components and their environment. By aligning Malinowski's ethnographic sensitivity with Bertalanffy's systemic abstraction, ISE achieves both depth and breadth, combining human-centred inquiry with structural coherence. This synthesis supports the development of systems that are not only technically robust but also attuned to the complex social ecologies in which they are embedded.

As ISE evolved into a distinct discipline, scholars recognised the need for interdisciplinary integration, combining perspectives from sociology, computer science, management, and engineering (Wangler and Backlund 2005). This convergence proved essential for addressing the ill-structured and dynamic problems characteristic of contemporary information systems. The accelerated development of digital technologies and the growing prevalence of data-driven decision-making have intensified the demand for a robust, adaptable theoretical framework that can accommodate rapid technological and societal change.

Today, the practical consequences of recent transformations demonstrate that Information Systems Engineering (ISE) faces an urgent need to revise its theoretical, methodological, and operational foundations to align with contemporary epistemic and technological realities. From a broader Systems Engineering perspective, this revision implies expanding beyond traditional hard systems approaches toward a more comprehensive integration of social, organisational, and human dimensions.

Contemporary systems are no longer closed technical constructs, but sociotechnical ecosystems characterised by complexity, uncertainty, and emergent behaviour. Consequently, the epistemological basis of Systems Engineering must recognise that valid knowledge includes not only structural models, but also contextual understanding shaped by multiple stakeholders. Methodologically, this requires combining quantitative and qualitative methods, integrating hard and soft systems approaches, and adopting iterative, adaptive processes suited to dynamic environments.

Operationally, systems must be designed for interoperability, resilience, and continuous evolution within interconnected networks. For Information Systems Engineering, this broader framework redefines its mission: it must encompass not only data and computation but also governance, ethics, and human agency. Ultimately, the renewal of ISE's foundations represents a necessary paradigm shift, ensuring its capacity to respond effectively to the complex and interdependent systems of the twenty-first century (Autili, et al. 2025).

## 2   Philosophy of Science and Computing

The philosophy of science, historically dedicated to the critical examination of scientific practice, has always sought to clarify how knowledge is constructed, validated, and communicated. In times of epistemic uncertainty and social transformation, its relevance becomes even greater. It reminds us that science is not only a collection of verified facts, but also a human enterprise structured by assumptions, values, and interpretive frameworks. Science, as discussed in the writings of Popper (Popper 1959), Lakatos (Lakatos 1976), Kuhn (Kuhn 1962), and Feyerabend (Feyerabend 1975), is not a construct immune to flaws or neutral in its assumptions. It is a human, collective, and situated activity, subject to revisions, controversies, and even paradigmatic ruptures.

The philosophy of science (Vallor 2022), historically devoted to examining how knowledge is constructed and validated, gains renewed significance in the age of computing. As thinkers have shown, science is not a neutral or infallible enterprise, but a human and collective activity shaped by assumptions, values, and interpretive frameworks. In parallel, computing, though a relatively young discipline, has emerged as a transformative force that redefines not only scientific practice but also our modes of reasoning and communication. The rise of artificial intelligence, predictive modelling, and data-driven automation extends the epistemological and ethical questions once posed to science into new domains, compelling us to reconsider the very nature of explanation, causality, and responsibility in an increasingly algorithmic world.

Computing, although a relatively young discipline, has become one of the central forces shaping contemporary society. It not only produces tools and algorithms but redefines our very modes of reasoning, communication, and interaction with the world. Technologies such as artificial intelligence, predictive models, and data-driven automation now influence decisions that impact every aspect of human life, including education, healthcare, governance, and the economy. This growing influence brings to the forefront philosophical questions about the nature of explanation, causality, representation, and ethical responsibility.

The rapid expansion of computational systems has also revealed deep tensions. The so-called black-box nature of many algorithms makes them opaque to scrutiny. Decisions are often delegated to systems whose internal logic remains inaccessible even to their designers (Bathaee 2018). This opacity gives rise to a crisis of trust: when results are accepted simply because they are produced by "intelligent" systems, science risks becoming an act of faith rather than of reason. The philosophy of science thus calls for renewed demands for epistemic transparency, as well as for criteria of validity, interpretability, and accountability that extend beyond efficiency or performance metrics.

From this perspective, philosophy provides essential conceptual tools to reorient computing toward reflexivity and responsibility. First, it requires science to remain self-critical, aware of its presuppositions, and open to revision. The scientific process should be understood not as the pursuit of infallible truth, but as a systematic effort to approach understanding through doubt, experimentation, and argumentation.

Second, the scientist or, in this case, the computer scientist, must be seen as an agent responsible not only for technical correctness but also for the epistemological, social, and ethical implications of his or her work. The design of algorithms, data models, and computational infrastructures involves choices that encode values: what is measured, what is ignored, who benefits, and who is excluded. These are not merely technical decisions but philosophical ones, grounded in conceptions of truth, fairness, and knowledge (Mittelstadt, et al. 2016).

Third, philosophy reminds us that scientific knowledge does not exist in isolation. It must be in dialogue with other forms of knowledge, including ethical, social, historical, and aesthetic perspectives, because science operates within a complex and dynamic society. This dialogue is especially important in computing, where technological artefacts directly influence human relationships, cognitive processes, and cultural formations.

Ultimately, the philosophy of science (Vallor 2022) invites computing to rediscover its moral and epistemic commitments. In a world increasingly governed by algorithms, the challenge is to ensure that computational systems serve truth, justice, and human flourishing rather than merely optimise or control outcomes. Bridging philosophy and computing is, therefore, not an abstract intellectual exercise, but an urgent necessity for the integrity of science and for the future of human society.

Information Systems Engineering (ISE) has always been situated at the intersection of technology and conceptual rigour. Its foundations, ontological, philosophical, semiotic, and mathematical, constitute a multidimensional framework for understanding how knowledge is represented and operationalised. The recent proliferation of LLMs introduces new questions to this framework: What kind of ontology do these models instantiate? How do they redefine the boundaries between symbol and meaning, data and knowledge, syntax, and semantics?

## 3  Ontological Foundations: From Conceptual Models to Generative Ontologies

Failures in information systems development are often linked to inadequate or flawed methodologies, particularly those involving conceptual modelling, which serves multiple purposes in the development process. Critiques of such methods commonly highlight their weak theoretical grounding, prompting several efforts to establish stronger foundations based on different reference disciplines. Although the relevance of ontology to data modeling was acknowledged as early as the 1950s, explicitly ontological approaches emerged only in the mid-1980s, when Wand and Weber, drawing on Mario Bunge's scientific ontology, developed what became known as the Bunge–Wand–Weber (BWW) ontology (Lukyanenko s.d.).

This approach has been criticised for its limitation to the material world, comprising physical objects with properties independent of human perception. Bunge's framework excludes human intentions, interpretations, and meanings, neglecting the "institutional reality" that encompasses socially constructed entities such as corporations, contracts, and transactions. Since these conceptual objects, central to organisational and informational contexts, have no representation within a purely material ontology, Bunge's framework is considered an inadequate foundation for conceptual modelling in organisational information systems.

Ontology, a central theme in this discourse, categorises the entities and relationships within information systems, thereby enhancing comprehension of the conceptual landscape. Philosophical considerations, such as epistemology and ethics, inform how knowledge is generated and utilised within these systems, enabling a deeper understanding of their functionality in social and organisational contexts. Semiotics further enriches this framework by examining the role of signs in communication and knowledge representation, ensuring clarity and precision in conveying information across different stakeholders.

In contrast to traditional approaches grounded in explicit ontological frameworks, LLMs operate within probabilistic language spaces, producing context-sensitive representations that lack predefined ontological commitments, raising the question of whether their observed success truly stems from an ability to reason over unstructured or semi-structured data, or rather from the effective internalization of linguistic patterns and contextual associations that function merely as implicit ontologies (Mai, H.T.; Chu, C.X.; Paulheim, H; 2025).

This operational mode necessitates a reconsideration of the very notion of conceptual modelling, as meaning becomes dynamic and emergent rather than fixed. The concept of a generative ontology, therefore, frames knowledge as continuously constructed through linguistic interaction, rather than statically represented. Such a perspective carries significant consequences for interoperability and knowledge integration within complex systems, challenging established assumptions about how conceptual structures are defined, communicated, and operationalised.

This tension characterises what Thomas Kuhn (Kuhn 1962) referred to as a paradigm crisis. According to the author, every scientific discipline evolves through long periods of normal science, during which a relatively stable conceptual framework guides practice, until the accumulation of anomalies and contradictions impedes the framework's functioning and imposes a paradigmatic rupture. In this sense, it can be stated with reasonable conviction that information systems have reached a critical point: the complexity of the social world and the integration of technology into everyday life have generated a set of problems that the traditional development model can no longer resolve.

In this context of epistemological and methodological disruption, Large Language Models (LLMs) emerge as a transformative response. Their generative capabilities offer a novel approach to ontology construction, particularly suited to the intricacies of sociotechnical systems. These systems are characterised by dense interrelations among human actors, institutional structures, and technological artefacts, all of which are documented across vast and heterogeneous textual corpora. LLMs possess the capacity to ingest, interpret, and synthesise this documentation, enabling the automated generation of ontologies that reflect both formal logic and the fluid semantics of human practice.

By bridging the semantic gap between human and machine understanding, LLMs facilitate the creation of ontological frameworks that are both contextually grounded and computationally tractable. This dual capacity addresses a core limitation of traditional information systems development: the inability to reconcile dynamic social realities with rigid technical schemas. Furthermore, the adaptability of LLMs enables continuous refinement in response to evolving system requirements, aligning with the agile, iterative nature of contemporary sociotechnical environments (Wu and Or 2025).

In this aspect, LLMs do not merely represent a technological innovation; they signify a paradigmatic shift in the epistemic foundations of information systems. One of the primary advantages of LLMs lies in their semantic richness and contextual awareness. Trained on expansive and varied corpora, these models can discern subtle linguistic nuances and complex relationships that are often essential for accurately modelling the multifaceted interactions within sociotechnical systems.

The generative capabilities of LLMs, combined with their proficiency in large-scale documentation analysis and semantic modelling, make them particularly well-suited for ontology generation in complex sociotechnical systems. Their integration into ontology engineering workflows promises to significantly enhance the responsiveness, accuracy, and relevance of knowledge representations in dynamic, interdisciplinary domains.

Ultimately, it is in the interplay between LLMs and human expertise that a genuine specification of quality in Information Systems emerges: the model contributes vast associative and inferential capacities across heterogeneous data spaces, while the human counterpart ensures semantic grounding, contextual interpretation, and ethical coherence. Together, they form a hybrid epistemic process in which computational scalability and human intentionality converge, enabling the construction of information systems that are not only technically consistent but also conceptually robust and socially meaningful.

Within this hybrid framework, Retrieval-Augmented Generation (RAG) architectures play a decisive role by integrating LLMs' generative capabilities with the precision and reliability of external knowledge retrieval. RAG introduces a dynamic coupling between symbolic repositories and probabilistic language models, allowing the system to access, filter, and reinterpret domain-specific information during the generation process. This coupling transforms LLMs from purely statistical predictors into adaptive cognitive instruments that can ground their outputs in verifiable and contextually relevant data sources.

Consequently, when embedded in human-AI collaboration, RAG enhances interpretability and accountability, two dimensions essential to the epistemic legitimacy of Information Systems Engineering. The resulting systems not only synthesise knowledge at scale but also support traceable reasoning, bridging the gap between automated inference and human understanding. In this sense, RAG operationalises the very principle that defines quality in information systems: the continuous mediation between data-driven intelligence and human discernment.

## 4  Philosophical and Epistemological Considerations

Systems thinking in computer science does not emerge in a conceptual void; it is rooted in a longstanding philosophical and scientific tradition that extends from the holistic reasoning of classical philosophy to the constructivist and cybernetic paradigms of modern computation (Rousso 2025). The epistemology of systems thinking in information systems emphasises the dynamic and relational nature of knowledge creation and its evolution. Knowledge is not seen as a mere collection of facts but as a construct shaped by interactions within systems. While its practical expressions appear in areas such as software engineering, information systems, and artificial intelligence, its theoretical structure reflects enduring questions about how complex systems can be represented, analysed, and influenced. At its core, systems thinking in computing engages with ontological issues concerning the nature of informational entities and processes, epistemological questions about how knowledge is formalised and validated in computational models, and ethical considerations regarding the consequences of automated decision-making. This philosophical grounding provides the conceptual depth needed to understand the evolving relationship among computation, representation, and reality.

Philosophically, LLMs challenge the correspondence model of truth that underlies most information systems. Their epistemology is not one of reference but of coherence, where truth is understood as internal consistency within vast corpora of text. Drawing on Heidegger's critique of technology (Heidegger 1977) and Floridi's philosophy of information (Floridi 2013), this paper examines how this new epistemic condition redefines the concept of "knowledge" in computational contexts.

Building on this philosophical foundation, the epistemic shift introduced by Large Language Models (LLMs) invites a re-examination of how knowledge is constituted, validated, and operationalised within computational systems. The coherence-based epistemology of LLMs diverges from the traditional correspondence model, which postulates that truth stems from a direct mapping between symbolic representations and external reality. Instead, LLMs generate meaning through probabilistic associations and semantic consistency across vast textual corpora, privileging internal relationality over external referentiality.

This reconfiguration has profound implications for the design and interpretation of information systems. Heidegger's view on technology suggests that we should be cautious not to treat people and the world around us as mere tools or resources to be exploited. He believed that modern technology encourages us to view everything, including ourselves, as something to be managed or optimised, rather than valued for its deeper meaning or purpose. In this light, LLMs challenge the instrumental rationality that dominates conventional system architectures, offering instead an emergent, contextual, and interpretive mode of knowledge production. Rather than encoding fixed truths, LLMs simulate understanding by dynamically aligning linguistic patterns, thereby emphasising the ontological conditions under which information becomes meaningful.

Luciano Floridi's philosophy of information further illuminates this transformation. His conceptualisation of knowledge as a function of semantic content, well-formedness, and truthfulness aligns with the generative logic of LLMs, although with critical tension. While Floridi emphasises the ethical and epistemic responsibilities of informational agents, LLMs operate within a framework where coherence may substitute for veracity, raising questions about the reliability and accountability of machine-generated knowledge. This tension highlights the need for new evaluative criteria that extend beyond binary notions of truth and falsehood, embracing instead a spectrum of epistemic virtues, such as relevance, transparency, and contextual fidelity.

In computational contexts, this epistemic condition redefines knowledge not as a static repository of facts but as a fluid, dialogical process shaped by interaction, interpretation, and iteration. LLMs exemplify this shift by enabling systems that learn, adapt, and respond to evolving discursive human-algorithm environments. They do not merely store or retrieve information; they participate in the construction of meaning, blurring the boundaries between data, knowledge, and understanding. Consequently, the integration of LLMs into information systems marks a paradigmatic departure from representationalism models, inviting a more reflexive and philosophically grounded approach to digital epistemology.

This epistemological shift, however, carries profound implications when LLMs interact with users whose beliefs diverge from established scientific consensus. In such interactions, the model's adaptive linguistic behaviour may unintentionally validate and reinforce non-empirical narratives, as coherence within dialogue is privileged over correspondence with objective reality. Without mechanisms of epistemic correction, the conversational process risks becoming a feedback loop of opinion, in which discursive plausibility substitutes for empirical verification. Semiotics offers a crucial lens for understanding LLMs' representational mechanisms. While Peirce's triadic model (sign–object–interpretant) (Liszka 1996) presupposes an interpretive subject, LLMs simulate interpretation algorithmically.

From a computational perspective, this phenomenon reveals a structural limitation in current generative architectures: their optimisation for linguistic consistency rather than epistemic integrity. Much like the phenomenon of collective hallucination in human cognition, an LLM aligns semantically with the user's expressed worldview, generating responses that maintain dialogical harmony but may drift from factual grounding. These dynamic positions the model within what Parmenides[1] described as the Path of Opinion, a domain of seemingly rational discourse detached from the pursuit of truth.

In the philosophy of information, this situation underscores the necessity of designing informational agents capable of epistemic self-regulation, that is, systems that can not only produce coherent linguistic outputs but also evaluate their correspondence with verified knowledge structures. Embedding such mechanisms requires a synthesis of systems thinking, epistemology, and computational ethics, ensuring that digital knowledge production remains anchored. The challenge, therefore, is not merely technical but conceptual: to engineer systems that preserve the fluidity of dialogical interaction while maintaining fidelity to the empirical world that sustains meaningful knowledge.

## 5      Ethical and Societal Implications

As Large Language Models (LLMs) and other AI systems become embedded in critical infrastructures, such as education, health, finance, and governance, their ontological opacity introduces new layers of ethical complexity. Unlike traditional automation, which is constrained by explicit procedural logic, generative models operate through probabilistic inference within vast semantic spaces. This opacity challenges the principles of accountability and transparency traditionally expected in corporate and institutional governance. Ethical evaluation, therefore, must extend beyond compliance and data privacy to encompass epistemic justice and the preservation of human interpretive autonomy. Moreover, it addresses the ethical dimension: if meaning is generated algorithmically, how should responsibility and authorship be rethought in the design of intelligent systems (Stahl 2023)?

From the standpoint of corporate social responsibility (CSR), organisations that adopt LLM-based solutions must acknowledge their dual role: as technological innovators and as moral agents within the sociotechnical ecosystem. Bias, misinformation, and the reinforcement of structural inequalities are not merely technical faults, but reflections of epistemic asymmetries encoded within the model's training data. Addressing these asymmetries requires more than corrective algorithms; it demands a framework guided by philosophical

---

[1] Greek pre-Socratic philosopher (530 BC to 460 BC)

From Symbol to Meaning    11clarity and moral prudence, ensuring that design choices align with human values and democratic accountability.

Transparency of Meaning – Systems must be designed to make their inferential processes, limitations, and epistemic boundaries intelligible to users and stakeholders. This transparency is not only procedural but semantic, recognising that meaning making is a human prerogative that cannot be fully delegated to machines (Russo, Schliesser and Wagemans 2024).

Plurality of Ontologies – Information systems must respect the diversity of cultural, linguistic, and interpretive frameworks through which human communities construct knowledge. The imposition of a single computational ontology risks homogenizing meaning and erasing epistemic diversity.

Preservation of Interpretive Autonomy – Automation should augment, not replace, human judgment. Decision-support systems must preserve users' capacity to interpret, deliberate, and dissent, thereby maintaining the ethical dimension of responsibility within human–machine interaction.

These principles reposition the debate on AI ethics from a compliance-oriented stance to an ontological ethics of responsibility. Rather than focusing exclusively on technical transparency or data governance, this approach seeks to articulate how AI can coexist with human interpretive freedom. Ultimately, the ethical maturity of AI-driven systems will depend less on the sophistication of their algorithms and more on organisations' commitment to designing, implementing, and regulating them in ways that respect the plurality of meanings and the moral agency of human beings (Gordon and Nyholm 2021).

The ethical dimension of automation in organisational contexts extends beyond technical implementation and encompasses profound implications for both humans and culture. Employees whose tasks are subject to automation often experience anxiety and resistance due to the perceived threat of job loss. Such reactions are not merely individual but are rooted in broader organisational cultures that may perceive automation as a disruption to established practices and professional identities. Consequently, any process of technological integration demands careful ethical reflection on how to mitigate these fears and ensure fair adaptation.

From an ethical standpoint, it is essential to design automation strategies that respect workers' dignity, provide opportunities for retraining, and promote transparent communication about the goals and consequences of technological change. The transition to automated processes must therefore be guided by principles of justice, empathy, and inclusion, recognising that resistance frequently arises from legitimate concerns about security and belonging. Addressing these issues requires not only managerial foresight but also moral responsibility for employees' psychological well-being and for preserving a healthy organisational culture.

## 6      Technical aspects

In recent years, the rapid development of large language models has transformed the way computational systems generate, interpret, and apply knowledge. These models have demonstrated remarkable capabilities in producing fluent and coherent text, solving complex problems, and adapting to a wide variety of domains. However, despite their versatility, such models remain limited by their intrinsic dependence on the data used during training.
Once deployed, they operate without direct access to external sources of information, relying exclusively on internalised representations that may not always reflect current, accurate, or contextually appropriate knowledge. This limitation often results in outputs that are plausible in form but factually inconsistent or disconnected from the specific demands of a given task. From a technical perspective, this issue arises because most large language models function in a closed inference regime, where parameters encode statistical correlations derived from pretraining corpora rather than dynamically validated facts. The embedding space, though rich in semantic associations, remains static after training, preventing the system from updating its representations in response to novel or evolving data. Moreover, the context window used during inference limits the model's capacity to retain or cross-reference extended informational contexts, resulting in partial or fragmented reasoning chains. These structural constraints, combined with the absence of retrieval or verification mechanisms, explain phenomena such as hallucination, domain drift, and semantic overgeneralization, where generated content prioritises linguistic plausibility over empirical correctness. Consequently, without integration of external retrieval, feedback loops, or grounding in structured knowledge bases, such models risk producing outputs that mimic understanding while lacking the epistemic robustness required for reliable decision-making in specialised or dynamically changing domains.

To overcome this challenge, recent research has proposed integrating retrieval-augmented generation as a means of grounding model outputs in relevant and verifiable information (Gao, et al. 2024), retrieval-augmented generation (RAG). In this framework, the generative process is coupled with an information retrieval mechanism that identifies and selects texts, examples, or solutions that bear contextual similarity to the problem at hand. By combining retrieved evidence with the model's generative reasoning, it becomes possible to produce responses that are both creative and factually aligned, thus bridging the gap between probabilistic inference and grounded knowledge.

This approach is grounded in two foundational observations. First, a wide range of programming and reasoning tasks exhibit profound structural commonalities, not only in their formulation but also in the cognitive and algorithmic strategies required to resolve them. Second, empirical evidence suggests that LLM performance can be substantially improved when they are provided with contextually relevant exemplars or analogous problem cases before generating a solution.

The concept of RAG represents one of the most significant advancements in the recent evolution of intelligent systems, particularly in their ability to connect symbolic knowledge with statistical language modelling.

By integrating external knowledge retrieval with generative modelling, RAG establishes a hybrid cognitive architecture: it combines the precision of information retrieval (IR) methods with the contextual adaptability of large language models (LLMs). In practical terms, when a query or prompt is introduced, the RAG framework first retrieves relevant information from a curated knowledge base or document repository, grounding the subsequent generation phase in verifiable and contextually aligned content. This mechanism mitigates one of the central limitations of purely generative systems, namely, their tendency toward hallucination or ungrounded inference, by constraining the model's reasoning within a corpus of externally validated data.

The integration of retrieval and generation produces a dynamic interaction between structured and unstructured representations. While the retrieval component ensures factual coherence and domain relevance, the generative component synthesises this information into coherent, human-like discourse. As a result, RAG does not merely recall or reproduce existing data; it reconstructs knowledge through contextual reasoning, enabling adaptive problem-solving across complex computational domains, such as programming, data analytics, scientific modelling, and decision support.

In this sense, RAG provides a promising framework for bridging the gap between abstract reasoning and concrete applications. It operationalises the transition from symbolic inference (anchored in explicit data) to probabilistic interpretation (characteristic of LLMs), thus aligning machine reasoning more closely with human cognitive processes. This convergence enables systems not only to generate linguistically coherent responses but also to reason effectively over retrieved evidence, leading to more robust, explainable, and semantically grounded outputs.

When a system retrieves analogous problems and their corresponding solutions, it effectively provides the model with a cognitive platform that supports reasoning through comparison and analogy. This process not only enhances performance but also contributes to epistemic transparency, as the sources of information become part of the reasoning context.

From a mathematical perspective, the efficacy of retrieval-augmented generation in problem-solving domains rests on principles from discrete mathematics, linear algebra, and probability theory. Structured problem instances can often be represented as graphs, matrices, or formal expressions, enabling the identification of isomorphisms or structural similarities between prior and current tasks. Algorithmic patterns, such as recurrence relations, combinatorial structures, or optimisation strategies, provide a formal substrate for retrieval mechanisms to operate on, guiding the selection of relevant analogues. Moreover, probabilistic models underlie the scoring and ranking of retrieved instances, allowing the system to weigh the likelihood that a particular prior solution will generalise effectively to a new context. This mathematical grounding ensures that the retrieval process is not merely associative but systematically aligned with the domain's formal properties, thereby reinforcing both the reliability and interpretability of the reasoning produced.

By aligning linguistic generation with structured retrieval, this approach represents a significant step toward a new generation of intelligent systems, ones that are not merely probabilistic but epistemically grounded and context aware. To address these limitations, retrieval-augmented generation has emerged as an effective method for connecting large language models with relevant external knowledge. In this approach, the model does not rely solely on its internal parameters but also utilises an information retrieval component to search a structured repository for texts like the current problem. The retrieved information is then combined with the model's own generative capacity, allowing it to produce answers that are more precise, coherent, and contextually grounded.

## 7      Considerations and Future Directions

The preceding discussions converge toward a shared understanding that the future of Information Systems Engineering (ISE) demands a reconfiguration of its conceptual, methodological, and ethical foundations. The convergence of intelligent technologies, societal transformation, and environmental urgency compels the discipline to articulate a renewed vision anchored in systemic coherence, epistemic responsibility, and sustainable innovation. The following considerations delineate the essential directions for future action.

ISE must advance beyond its traditional technological orientation to incorporate sustainability as a fundamental principle, rather than an external constraint. The forthcoming generation of engineered systems should be self-adaptive, energy-efficient, and resilient, capable of operating within complex ecological and socio-technical contexts. This evolution requires embedding sustainability metrics into the earliest stages of design, ensuring that resource optimisation and environmental stewardship become intrinsic components of engineering decisions. The discipline must therefore align its methods with global imperatives of sustainability and ethical responsibility, redefining success not only in terms of efficiency but also in terms of long-term societal benefits.

The expansion of societal demands, ranging from digital inclusion to equitable access to information, necessitates a systemic reconceptualisation of design goals. Future systems must evolve as dynamic agents capable of co-adaptation with changing social structures and values. This entails synchronising individual system life cycles with collective challenges such as public health, education, and social equity. The methodological response should foster interdisciplinary collaboration, bridging engineering, social sciences, and ethics to address ill-structured problems that defy traditional technical boundaries.

Artificial intelligence and machine learning are not merely instruments of automation but epistemic agents[2] that reshape the way systems are conceived and governed. Their deployment introduces profound ethical challenges concerning autonomy, accountability, and the transformation of human labour. The future practice of ISE must institutionalise ethical risk assessment as a fundamental stage in the design process, ensuring that innovations contribute positively to human welfare. Governance mechanisms should promote transparency, explicability, and fairness in automated decision-making, counterbalancing the asymmetries that emerging technologies may exacerbate.

Global technological interdependence renders traditional governance frameworks inadequate. The future of systems engineering lies in constructing polycentric governance models, which integrate governments, academia, civil society, and the private sector in a shared responsibility for technological evolution. Such models must foster open dialogue and cooperative regulation to manage uncertainty, mitigate systemic risk, and uphold the primacy of the public good. Governance, in this sense, becomes an adaptive process that co-evolves with technological and societal transformations.

Finally, ISE must cultivate a reflexive stance, acknowledging that technological systems are not neutral artefacts but active participants in shaping human knowledge and social order. This recognition demands a continuous re-examination of the epistemological assumptions underlying engineering practice. The discipline's renewal thus depends on its capacity to integrate philosophical reflection, ethical foresight, and empirical rigour into a unified framework of responsible innovation.

## 8    Conclusion

The reflections developed in this paper converge on a central claim: Information Systems Engineering (ISE) must confront the epistemic and ontological consequences of the computational transformations it now embodies. The emergence of Large Language Models (LLMs) represents far more than a technological innovation; it signals a profound shift in how knowledge, meaning, and representation are conceived within information systems. Traditional engineering paradigms, grounded in stable requirements, deterministic logic, and fixed ontological schemas, are increasingly unable to capture the fluid, dialogical, and context-sensitive nature of contemporary digital ecosystems. The convergence of philosophy, ontology, and systems thinking reveals that the current evolution of computational technologies cannot be adequately understood within traditional paradigms. These models challenge the correspondence view of truth, reconfigure the processes of meaning-making, and redefine the very conditions under which knowledge is produced, validated, and applied in digital environments.

---

[2] The expression *epistemic agents* designates entities that can engage in epistemic activities, that is, activities related to the acquisition, production, justification, representation, and communication of knowledge. The concept is rooted in Epistemology, where it is used to characterise those who can hold beliefs, evaluate evidence, and revise their cognitive states in light of new information.

    The epistemic and ethical implications of this transformation demand a reorientation of the discipline toward reflexivity, transparency, and accountability. Information systems can no longer be conceived merely as technical artefacts; they are epistemic agents that participate in the construction of social reality. Their design, therefore, entails philosophical responsibility. The interplay between ontological clarity and computational generativity must be consciously managed to ensure that the systems we engineer continue to serve human values, intellectual integrity, and democratic participation.

Philosophy of science provides the critical framework for this reorientation. It reminds us that science, and by extension, computing, is a human endeavour governed not solely by logic and efficiency, but by interpretive frameworks, ethical commitments, and social contexts. Likewise, ontology supplies the formal rigour required to structure knowledge, while semiotics mediates between meaning and representation. These perspectives, when integrated within the engineering of information systems, foster an understanding of computation that is not reductionist but systemic, not instrumental but interpretive.

Ethically, this synthesis reaffirms the primacy of human interpretive autonomy. In a landscape increasingly dominated by algorithmic mediation, the preservation of human judgment, plurality of ontologies, and transparency of meaning become moral imperatives. The responsibility of researchers, engineers, and organisations extends beyond innovation; it encompasses cultivating epistemic justice and preventing informational asymmetries that undermine public trust and intellectual diversity.

**Acknowledgments.** This study was partially funded by the Brazilian *Conselho Nacional de Desenvolvimento Científico e Tecnológico* (CNPq) by grants 402.086/2023-6 and 306.695/2022-7.

**Disclosure of Interests.** The author has no competing interests to declare that are relevant to the content of this article. During the drafting of this paper, LLMs were utilised to assist in editing and condensing partial drafts written by the author. The author then further edited and refined the text with Grammarly. Such use and this acknowledgement adhere to the ethical guidelines for the use of generative AI in academic research. The author has developed the work entirely, which has been thoroughly vetted for accuracy, and assumes responsibility for the integrity of their contributions.

**References**

Autili, Marco , Martina De Sanctis, Paola Inverardi, and Patrizio Pelliccione. "Engineering Digital Systems for Humanity: A Research Roadmap." *ACM Transactions on Software Engineering and Methodology*, 26 May 2025, 5 ed.: 1 - 33.

Bathaee, Yavar . "The Artificial Intelligence Black Box and the Failure of Intent and Causation." *Harvard Journal of Law & Technology*, Spring 2018, 2 ed.: 890-938.

Feyerabend, Paul. *Against Method: Outline of an Anarchistic Theory of Knowledge.* London: New Left Books, 1975.

Floridi, Luciano. *The Ethics of Information.* Great Clarendon Street, Oxford: Oxford University Press, 2013.

Gao, Yunfan , et al. "Retrieval-augmented generation for large language models: A survey." *arXiv preprint.* Mar 27, 2024.

Gizzardi, Giancarlo. *Ontological Foundation for Structural Conceptual Models.* Twente,: Center for Telematics and Information Technology, University of Twente, 2005.

Gordon, John-Stewart , and Sven Nyholm. "Ethics of Artificial Intelligence." *Internet Encyclopedia of Philosophy.* Feb. 2021. https://iep.utm.edu/ethic-ai/ (accessed 10 19, 2025).

Heidegger, Martin. "The Question Concerning Technology." In *Basic Writings*, edited by David Farell Krell, 287-317. New York: Harper & Row, 1977.

Kuhn, T. S. *The Structure of Scientific Revolutions.* Chicago: University of Chicago Press, 1962.

Lakatos, Imre. *Proofs and Refutations.* Cambridge: Cambridge University Press, 1976.

Liszka, James Jakób. *A General Introduction to the Semeiotic of Charles Sanders Peirce.* Bloomington and Indianapolis: Indiana University Press, 1996.

Lukyanenko, R., Storey, V.C. & Pastor, O. "Foundations of information technology based on Bunge's systemist philosophy of reality." *Softw Syst Model*, n.d.: 921–938.

Mai, H.T.; Chu, C.X.; Paulheim, H;. "Do LLMs Really Adapt to Domains? An Ontology Learning Perspective." Edited by G., et al. Demartini. *Lecture Notes in Computer Science, The Semantic Web – ISWC 2024.* Springer Nature Switzerland, 2025. 126-143.

Malinowski, Bronislaw. "Practical Anthropology." *Journal of the International African Institute*, Jan. 1929: 22-38.

Mittelstadt, Brent Daniel , Patrick Allo, Mariarosaria Taddeo, Sandra Wachter, and Luciano Floridi. "The ethics of algorithms: Mapping the debate." *Big Data & Society*, 2016, July–December ed.: 1 - 21.

Popper, K. *The logic of scientific discovery.* New York, NY: Basic Books, 1959.

"Chapter 3 – Philosophical Foundations of Systems Thinking." In *Systems Thinking*, by Dotan Rousso. Calgary, AB: Southern Alberta Institute of Technology, 2025.

Russo, Frederica, Eric Schliesser, and Jean Wagemans. "Connecting ethics and epistemology of AI." *AI & Society: Knowledge, Culture and Communication*, 2024: 1585–1603.

Stahl , Bernd Carsten . "Embedding responsibility in intelligent systems: from AI ethics to responsible AI ecosystems." *Scientific Reports* , 18 May 2023.

Vallor, Shannon, ed. *The Oxford Handbook of Philosophy of Technology.* New York, NY: Oxford University Press, 2022.

Von Bertalanffy, Ludwig. "The history and status of general systems theory." *Academy of management journal*, 1972: 407-426.

Wangler, Benkt, and Alexander Backlund. "Information systems engineering: What is it?" *CAiSE 2005 Workshops.* Porto, Portugal: FEUP Edições, 2005. 427-437.

Wu, Ju , and Calvin K. L. Or. "Position Paper: Towards Open Complex Human–AI Agents Collaboration Systems for Problem Solving and Knowledge Management." *arXiv.* 9 Oct 2025.

Yearworth, Mike . "The theoretical foundation(s) for Systems Engineering?" *Systems Research and Behavioral Science*, Jan. 2020: 1-4.